\documentclass[10pt,twocolumn]{IEEEtran}
\usepackage{graphicx}
\usepackage{amsmath}
\usepackage{amssymb}
\usepackage{eso-pic}
\usepackage[noadjust]{cite}
\usepackage{float}
\usepackage{color}
\usepackage{url}
\usepackage{graphicx}
\usepackage{amsfonts}
\usepackage{float}
\usepackage{color}
\usepackage{cite}
\usepackage{epstopdf}
\usepackage{bm}
\usepackage{mathtools}
\usepackage[algoruled,resetcount,linesnumbered]{algorithm2e}
\usepackage{array}
\usepackage{multirow}
\usepackage{multicol}
\usepackage{mathtools}
\usepackage{tablefootnote}
\usepackage{caption}
\usepackage{subcaption}
\newcolumntype{P}[1]{>{\centering\arraybackslash}p{#1}}

\usepackage{lipsum}
\usepackage{fancyhdr} 
\graphicspath{ {fig/} } 

\makeatletter

\def\ps@IEEEtitlepagestyle{%
    \def\@oddfoot{\mycopyrightnotice}%
    \def\@evenfoot{}%
    }

\def\mycopyrightnotice{%
    {\footnotesize  978-1-7281-6861-6/20/\$31.00  \textcopyright2020 IEEE\hfill}
    \gdef\mycopyrightnotice{}
}

\newcommand\AtPageUpperMyright[1]{\AtPageUpperLeft{%
 \put(\LenToUnit{0.5\paperwidth},\LenToUnit{-1cm}){%
     \parbox{0.5\textwidth}{\raggedleft\fontsize{9}{11}\selectfont #1}}%
 }}%
\newcommand{\conf}[1]{%
\AddToShipoutPictureBG*{%
\AtPageUpperMyright{#1}
}
}

\begin{document}

\title{Localization with  Deep Neural Networks using mmWave Ray Tracing Simulations}
\author{Udita Bhattacherjee, Chethan Kumar Anjinappa, LoyCurtis Smith,  Ender Ozturk, and Ismail Guvenc\\
\IEEEauthorblockA{Department of Electrical and Computer Engineering, North Carolina State University, Raleigh, NC
}\\
Email: {\tt \{ubhatta,canjina,ljsmith9,eozturk2,iguvenc\}@ncsu.edu}}
\conf{IEEE SoutheastCon2020}
\maketitle

\begin{abstract}

 The world is moving towards faster data transformation with more efficient localization of a user being the preliminary requirement. This work investigates the use of a deep learning technique for wireless localization, considering both millimeter-wave (mmWave) and sub-6 GHz frequencies. The capability of learning a new neural network model makes the localization process easier and faster. In this study, a Deep Neural Network (DNN) was used to localize User Equipment (UE) in two static scenarios. We propose two different methods to train a neural network, one using channel parameters (features) and another using a channel response vector, and compare their performances using preliminary computer simulations. We observe that the former approach produces high localization accuracy: considering that all of the users have a fixed number of multipath components (MPCs), this method is reliant on the number of MPCs. On the other hand, the latter approach is independent of the MPCs, but it performs relatively poorly compared to the first approach.  
\end{abstract}


\section{Introduction}
In the upcoming 5th generation (5G) wireless communication networks, one of the most promising enhancements will be larger data rates with increased coverage, which requires faster beamforming in a given direction to maintain uninterrupted communication. To accomplish this, a base station (BS) must know a user equipment’s (UE) location within the network. The process of determining the location of a given UE within a particular area is called localization. Capability of localizing a UE can further be leveraged to provide location-based services by the cellular network. Thus, the process of localization is highly necessary in wireless communication.

In wireless networks, there exist many localization algorithms. In \cite{ni2017accurate}, the authors perform localization with the help of signaling data like Reference Signal Received Power (RSRP) and timing advance. In \cite{coluccia2019review}, authors compare different advanced algorithms, such as localization with hybrid Received Signal Strength (RSS) and Angle of Arrival (AOA), projection onto convex set, multi-hop methods, etc.  In \cite{guvenc2009survey}, the authors explore different Time-of-Arrival (TOA) based algorithms for localization. In general, these algorithms use channel parameters such as AOA, TOA, RSS as well as various channel statistics, derived from the channel parameters, to perform accurate localization. These procedures often involve time-consuming and complex operations, such as the least square methods mentioned in \cite{coluccia2019review}. A geometry based perspective to improve localization performance using the non-line of sight (NLOS) paths is explored in \cite{ruble2018wireless,ruble2018massive}.

Researchers considered NLOS components as source of distortion in earlier studies \cite{Direct_loc}. 
However, the NLOS components increase the channel sparsity as few MPCs can be received with significant RSS, hence provide additional information about the location of a UE \cite{NLOS_imp1}. In \cite{NLOS_imp2}, it is mathematically demonstrated that NLOS components are the most informative ones in case of narrow beams (e.g. mmWaves).  
The most popular features used to predict the location in the literature are AOA, TOA, and RSS. The authors in \cite{TOA-AOA} provides the performance analysis of localization using these features. They observed different combinations of these features and found the result to be better for the combination of TOA and AOA, but not much reliable with RSS. The numerical analysis of these features using Monte-Carlo simulation can be found~in~\cite{Monte-Carlo}.

The aforementioned works tackle the localization problem from a system modeling, signal processing, and even a geometry-based perspective. In this study, we leverage machine learning (ML) aiming to improve the localization process, in terms of low run-time complexity i.e. lower computation time without sacrificing the accuracy. The ML algorithms are capable of learning complicated functions if provided enough training samples. It is used extensively in the field of wireless communication for various tasks, such as the prediction of the Angle of Departure (AoD) channel feature from AOA \cite{navabi2018predicting}, predicting channel characteristics of a massive multiple-input multiple-output (MIMO) system \cite{bai2018predicting}, classification of different types of Unmanned Aerial Vehicles (UAVs) \cite{ezuma2019micro}, etc. ML is being used in localization research as well. In \cite{comiter2017data}, the authors observe the effects of a BS or an eNodeB (eNB) on node localization. In this study, a deep neural network (DNN) is proposed to locate user nodes in a mmWave network. However to the best of our knowledge, there is no study in the literature that compares mmWave and sub-6~GHz bands using ML techniques and explore the effect of using different channel parameters.
 
We use a supervised ML technique in this paper and pose the localization problem as a regression problem. In supervised learning, it is assumed that one has access to a set of learning features, measured over several observations, and an outcome variable ( i.e.the UE location in this paper), which is also known as the label or the target. The learning features can either be the combination of raw channel parameters, such as AOA, TOA, and RSS, or the channel response vector; see Section II for more details. The training data for the DNN was generated by Remcom's Wireless Insite~$\textregistered$, a ray-tracing simulator. Our preliminary results show that the proposed localization technique gives high accuracy considering a high signal-to-interference ratio (SNR) regime. Our future work includes studying the performance trade-offs in various different environments and SNRs, and exploring the effect of beam forming on the localization accuracy.

The rest of this paper is organized as follows. Section~II introduces the system model and problem formulation, including two different approaches for generating inputs for the DNN. Section~III introduces the proposed DNN technique. Section~IV provides our preliminary simulation results, Section~V discusses about the future direction of this work, and Section~VI concludes the paper.  



\section{System Model and Problem Formulation}
In this section, we describe how the channel features and location information provided by the ray-tracing model are transformed into input features and output labels for training a DNN at the BS side. In this process, we will mention two different approaches that we followed with relevant details. 

As described briefly in Section I, the training data for the DNN was obtained from ray-tracing simulations. The ray-tracing simulator generates channel parameters that capture the dependence of the environment geometry and transmitter/receiver locations, which are crucial for the ML applications. 
The choice of a dataset depends on our approach to designing the DNN, which is discussed up next.

\subsection{Approach 1: Utilization of the Channel Parameters}
In this approach, we use raw channel parameters observed at the BS, provided by the simulator i.e. AOA, TOA, and RSS, to train the DNN model in order to predict the location of the users. The number of inputs to the model depends on the selected number of MPCs and the number of features considered. For simplicity, we fixed the number of MPCs to three, and we used three different combinations of channel parameters to feed the model. Details of the DNN architecture for Approach 1 are as follows:
\begin{itemize}
    \item \textbf{Input:}
The number of inputs will vary for each combination of features - it will be the product of the number of MPCs and number of features.
\begin{itemize}
    \item Only AOA (3 inputs)
    \item Only AOA and RSS (6 inputs)
    \item Only AOA, RSS and TOA (9 inputs)
\end{itemize}
\end{itemize}

\begin{itemize}
\item \textbf{Output:}
We consider the output to be the x and y coordinates of a user's location. We assume all the users are on the same plane, thus we ignore the z coordinates.
\end{itemize}
Details of the trained DNN model are further discussed in Section III.

Intuitively, more MPCs will provide more information to the DNN to be trained more efficiently, but it will increase the training computational complexity and likelihood of overfitting as well. In this trade-off, we defined the number of MPCs as three to have the balance between accuracy and complexity.

\begin{figure}[!t]
    \centerline{
    \includegraphics[scale=0.45]{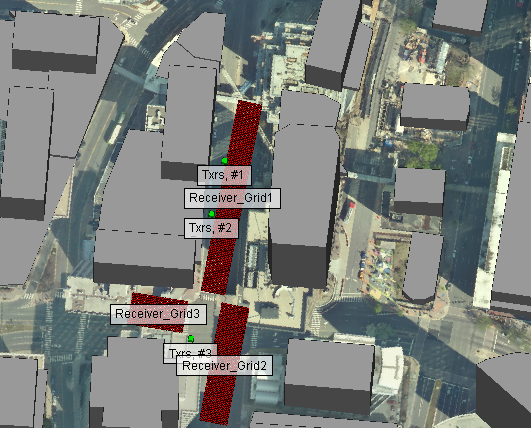}}
    \caption{Approach 1: Ray tracing scenario in North Moore Street, Rosslyn, Virginia.}
    \label{fig:approach1}
\end{figure}

Fig.~\ref{fig:approach1} shows a scenario where a UE communicates with BSs. We set up multiple UE grids (red path) and BS nodes (green point) within an urban environment. All of the nodes use half-wave dipole antennas at a height of 10 m from the ground and transmits at a power level of 0 dBm. They are excited with two different frequencies, 5~GHz and 28~GHz, with 100~MHz and 500~MHz bandwidths, respectively.

Specifically, for this study, \textit{Txrs} $\#1$ is the transmitter for which we observe the channel parameters for both receiver grid 2 and 3. All of the users in receiver grid 2 are in Line of Sight (LOS), and the users in receiver grid 3 are in Non-Line of Sight (NLOS). The UEs are spaced 1~meter apart from each other, and there are a total of 1530 receiver points, 990~in receiver grid 2, and  540~in receiver grid 3. The UE locations, within their respective receiver grid, serve as the expected output of the DNN. 


\textbf{Drawback of this approach:} 
In this approach, we only consider a fixed number of MPCs which might not be a realistic assumption. This is because the actual number of MPCs vary in practice for each user due to the dynamic environment and scatterers. To make the model more flexible and independent of the number of MPCs, we took another approach, which is described below.

\subsection{Approach 2: Utilization of the Channel Response Vector}
In this approach, we considered a channel response vector as the input to the DNN model. Namely, we took the number of antennas as inputs, which makes the system robust against the number of varying MPCs. For simplicity, we assumed a single antenna on the UE and set the number of antennas on the BS to 10. Details of the DNN architecture for Approach 2 is as follows: 
\begin{itemize}
    \item \textbf{Input:}
The number of inputs will be the product of antennas at the BS and UE.
\begin{itemize}
    \item We pass the absolute value of the channel response vector as the input. The absolute value of the channel impulse response was calculated using functions provided in  \cite{alkhateeb2019deepmimo}. \footnote{https://www.deepmimo.net/}
\end{itemize}
\end{itemize}

\begin{itemize}
\item \textbf{Output:}
We consider the output to be the x and y coordinates of a user's location. We assume all the users are on the same plane, thus we ignore the z coordinates.
\end{itemize}

\begin{figure}[!t]
    \centerline{
    \includegraphics[scale=0.6]{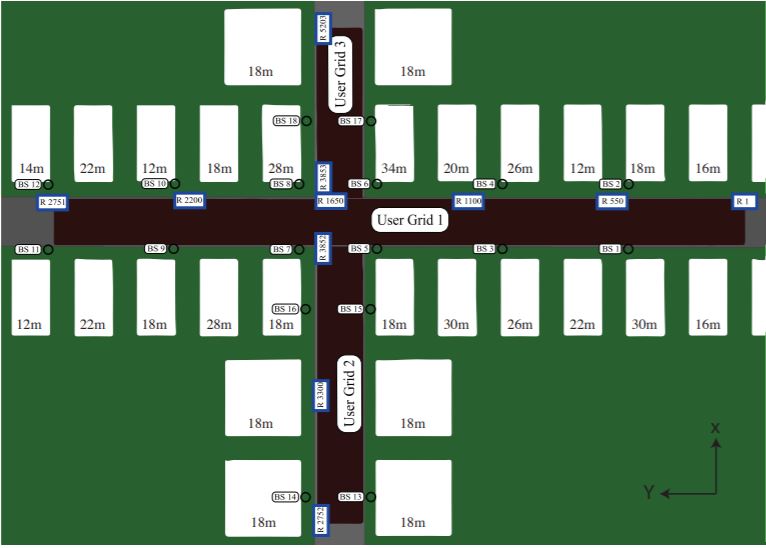}}
    \caption{Approach 2: Ray tracing scenario from DeepMIMO dataset.}
    \label{fig:approach2}
\end{figure}

We used a scenario where UEs communicate with different BSs as shown in Fig. \ref{fig:approach2}. The Deep-MIMO dataset was generated for ML researchers. For our DNN, we have considered BS3 as the transmitter and the users (1000 - 1025 rows) from User grid 1 as receivers, which is 185 users in total. For each user, the channel response is generated as follows:
\begin{equation}\label{Equation}
    \mathbf{h}_k^{b,u} = \sum_{l=1}^L \sqrt{\frac{\rho_l}{K}} e^{j\left(\vartheta_l^{b,u} + \frac{2\pi k}{K} \tau_l^{b,u}B  \right)}\mathbf{a}(\phi_{az}^{b,u},\phi_{el}^{b,u})~,
\end{equation}
where $h_k^{b,u}$ is the channel response vector at BS $b$ from the UE $u$ for subcarrier $k$. The $\rho_l, \vartheta_l,\tau_l,\phi_{az},\phi_{el}$ are the RSS, Doppler frequency, TOA, AOA (azimuthal and elevation), respectively. The number of MPCs is denoted by~$L$.

\section{Deep Neural Network Preliminaries}
In this section, we discuss some preliminaries and the overall architecture of the considered DNN.

\subsection{Input Features}

\begin{figure}[t]
    \centering
    \includegraphics[scale=0.5]{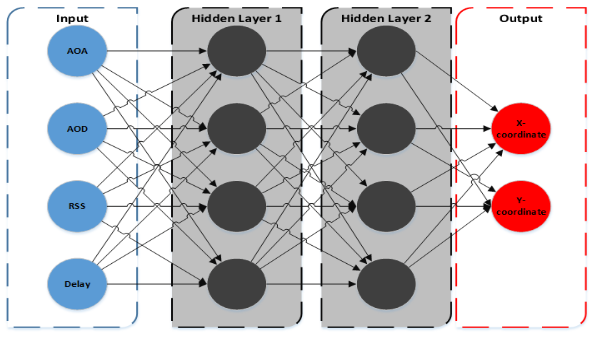}
    \caption{Illustration of DNN for approach 1.}
    \label{fig:DNN}
\end{figure}

As discussed earlier, the input features depend on the approach. For both approaches, we use the combination mentioned in Section~II with appropriate normalization to the data set. An illustration of Approach~1 is shown in Fig.~\ref{fig:DNN}.

\subsection{Output Labels}
The Cartesian coordinates x and y of the location are considered as the output labels. The z coordinates are ignored since all the users are in same plane.

\subsection{Hidden Layers and Hyper-parameters}
Two hidden layers, as shown in the Fig. \ref{fig:DNN}, are considered in this model. The other hyper-parameters, such as number of nodes in each hidden layer, learning rate and activation function, are optimized using the Bayesian optimization technique by calculating the mean square error (MSE). The grid of hyper-parameters used while training the DNN model are shown in Table \ref{tab:T_FI_0}.

\begin{table}[!h]
\caption{Hyper-parameters and their range.}
\begin{tabular}{|P{3.5cm}|P{4.25cm}|}
\hline
\textbf{} & \textbf{Range} \\
\hline
\textbf{Number of Nodes }  & 4 to 50 \\
\hline
\textbf{Learning Rate}  & 1e-3 to 1e-1 \\
\hline
\textbf{Activation Functions}  &  tansig, logsig, purelin (linear), poslin (positive linear), radial basis (radbas) \\
\hline
\end{tabular}\label{tab:T_FI_0}
\vspace{-0.5cm}
\end{table}

\subsection{Objective Function}
The objective of a supervised learning process is to minimize a loss function. An example of a loss function is the binary cross-entropy in classification and the mean square error (or the quadratic loss) in regression type problems. We used the mean square error as the loss function in this study.  We predict the location as a learning-based optimization problem, as shown in \eqref{Eq1}. The goal of the optimization problem is to learn the mapping $\mathcal{F}$, such that the MSE between the known output and the estimated output is minimized: 
\begin{equation}
\min_{\mathcal{F}} ||\mathcal{F}(\text{features}) - \text{output labels} ||^2~.
\label{Eq2}
\end{equation}
The DNN model has been designed using MATLAB’s deep learning toolbox. Given enough training data, the DNN can be trained well enough to learn complicated functions using the back-propagation algorithm. 

\begin{table*}[t!]
\caption{Frequency bands used in ray tracing simulations.}
\begin{tabular}{|P{2cm}|P{2cm}|P{3cm}|P{2cm}|P{2cm}|P{2cm}|P{2cm}|}
\hline
\textbf{Frequency} & \textbf{Type of path} & \textbf{Inputs} &  \textbf{Number of nodes in hidden layer 1} &  \textbf{Number of nodes in hidden layer 2} & \textbf{Learning Rate} &  \textbf{Activation Function}\\
\hline
\multirow{6}{*}{5~GHz} & \multirow{3}{*}{LOS} & AOA & 40 & 50 & 0.9078 & logsig\\
\cline{3-7}
& & AOA + RSS & 8 & 25 & 0.0010027 & logsig\\
\cline{3-7}
&  & \footnotesize{AOA + RSS + TOA} & 4 & 50 & 0.0011179 & tansig\\
\cline{2-7}
& \multirow{3}{*}{NLOS} & AOA & 40 & 38 & 0.40981 & radbas\\
\cline{3-7}
&  & AOA + RSS & 42 & 50 & 0.1957 & tansig\\
\cline{3-7}
&  & AOA + RSS + TOA & 29 & 31 & 0.93405 & tansig\\
\cline{2-7}
\hline
\multirow{6}{*}{28~GHz} & \multirow{3}{*}{LOS} & AOA & 50 & 41 & 0.97691 & tansig\\
\cline{3-7}
& & AOA + RSS & 8 & 46 & 0.0012802 & tansig\\
\cline{3-7}
&  & AOA + RSS + TOA & 15 & 46 & 0.8885 & logsig\\
\cline{2-7}
& \multirow{3}{*}{NLOS} & AOA & 34 & 50 & 0.0010755 & radbas\\
\cline{3-7}
&  & AOA + RSS & 9 & 50 & 0.9795 & tansig\\
\cline{3-7}
&  & AOA + RSS + TOA & 35 & 35 & 0.32042 & logsig\\
\cline{1-7}
\end{tabular}\label{tab:T_FI}
\end{table*}

\begin{figure*}[t!]
\centering
\begin{subfigure}{.45\textwidth}
  \centerline{\includegraphics[scale=0.5]{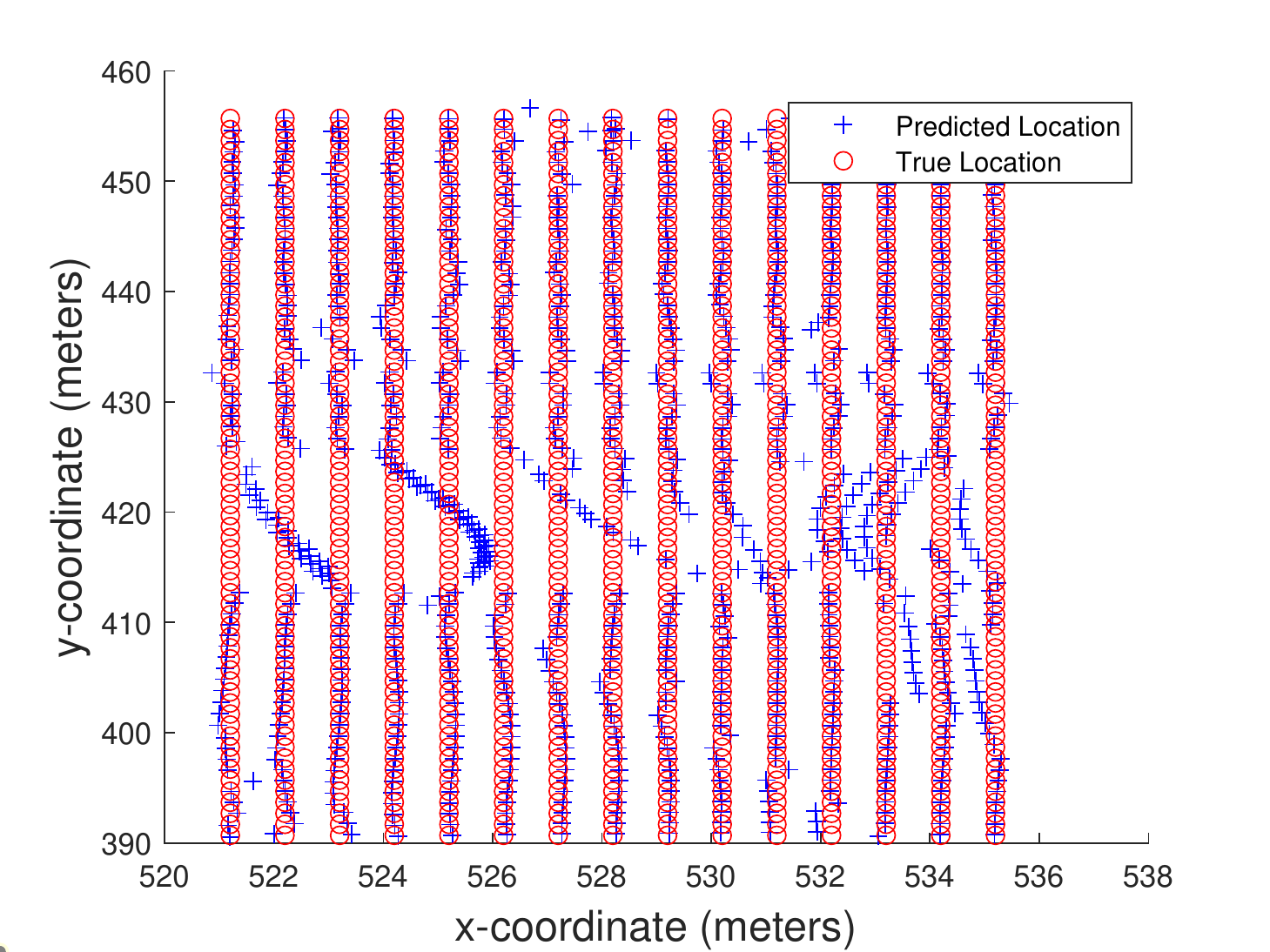} }
  \caption{Actual and predicted location using only AOA.}
  \label{fig:sub-first}
\end{subfigure}
\begin{subfigure}{.45\textwidth}
  \centerline{\includegraphics[scale=0.5]{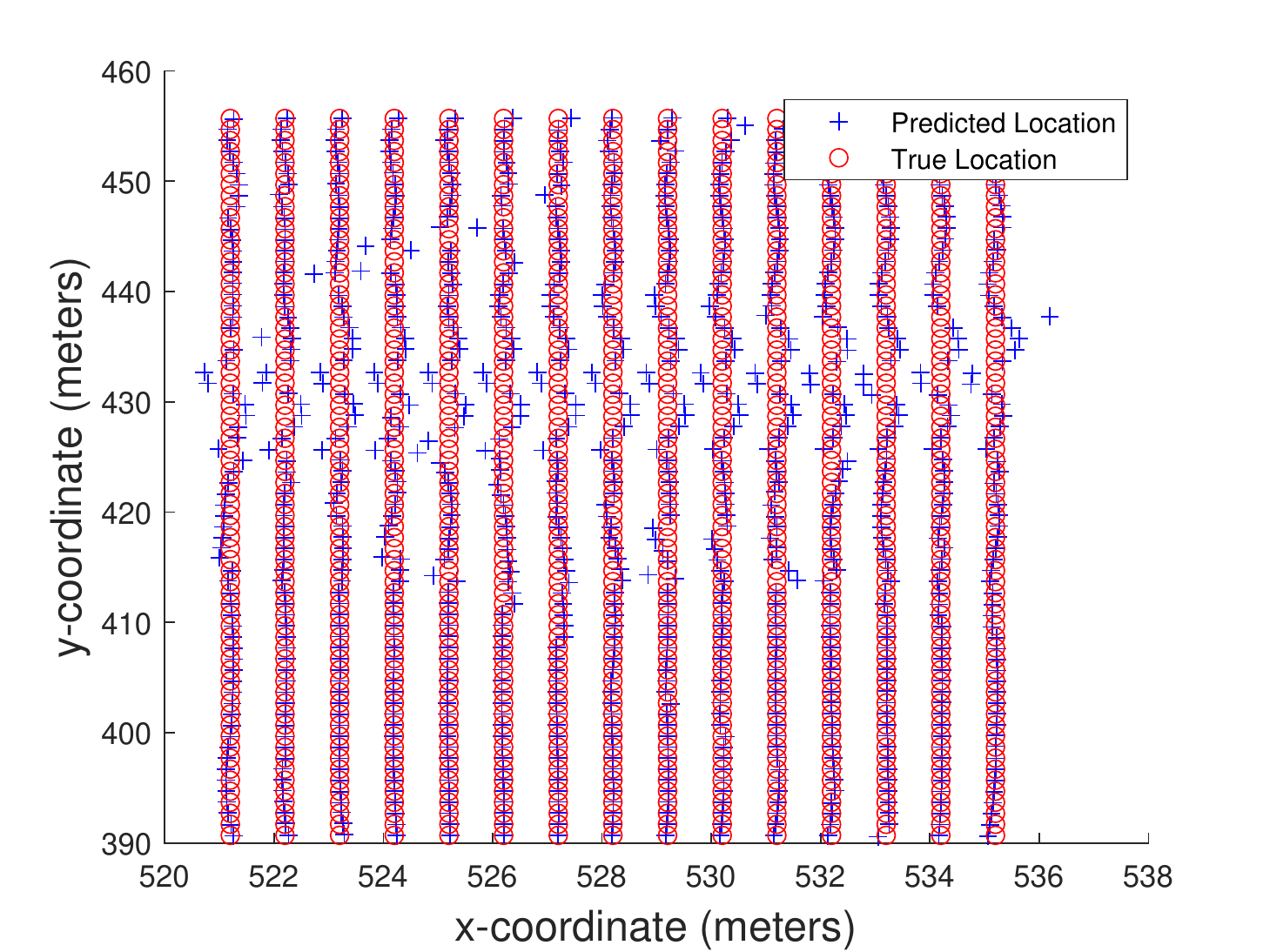}}
  \caption{Actual and predicted location using AOA and RSS.}
  \label{fig:sub-second}
\end{subfigure}
\begin{subfigure}{.45\textwidth}
  \centerline{\includegraphics[scale=0.5]{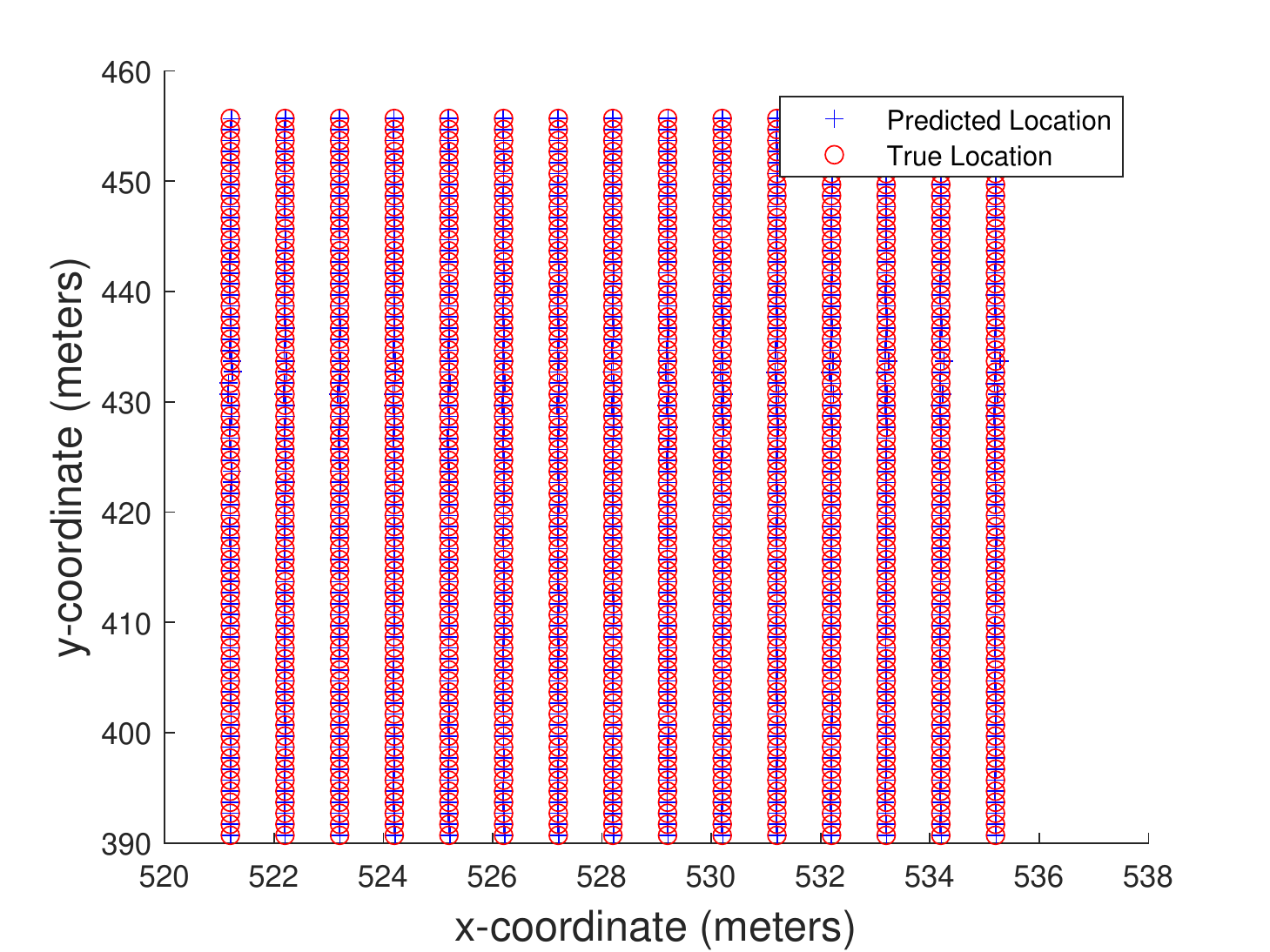} }
  \caption{Actual and predicted location using AOA, RSS and TOA.}
  \label{fig:sub-third}
\end{subfigure}
\begin{subfigure}{.45\textwidth}
  \centerline{\includegraphics[scale=0.5]{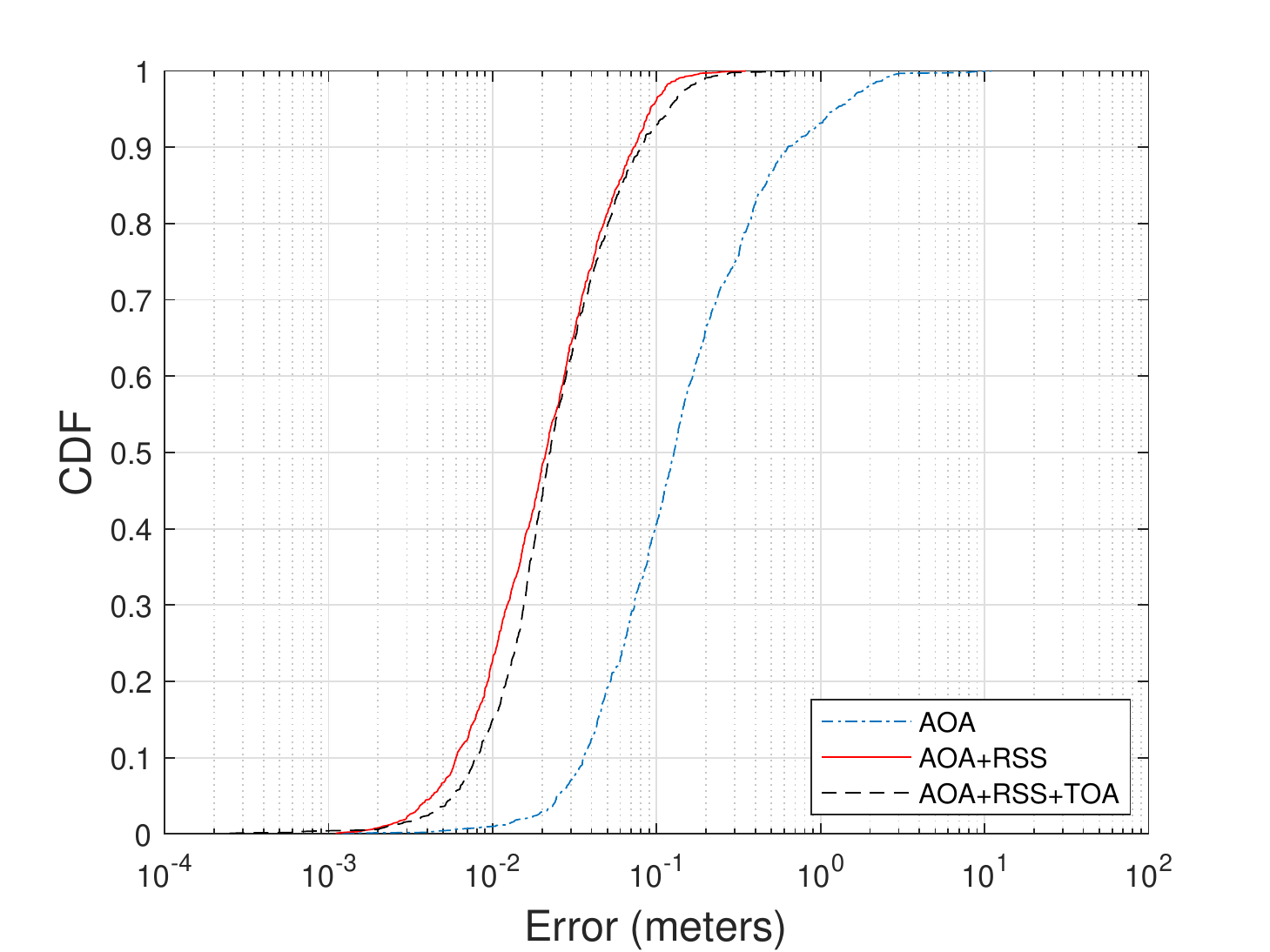} }
  \caption{CDF plot.}
  \label{fig:sub-fourth}
\end{subfigure}
\caption{Location and CDF plot at 5~GHz (LOS).}
\label{fig:5GHz_LOS}
\end{figure*}

\section{Simulation Results}
In this section, we discuss the simulation setup and the results obtained. For Approach~1, we divided the  localization problem into two cases, LOS-based and NLOS-based localization. Apriori, we know that all of the UEs in receiver grid 2 are in LOS and thus, possess a strong direct LOS signal component. Whereas, the UEs in receiver grid 3 are in NLOS. The total number of UEs for the LOS and NLOS cases are 990 and 540 user points, respectively. We performed the simulations at both 5~GHz and 28~GHz bands. For Approach~2, the considered region is located within the LOS area and has a total of 185 users.

For both approaches, the entire dataset is used for both training and testing purposes. The final hyper-parameters are obtained by performing Bayesian optimization. The trained model is used for testing the accuracy of proposed localization approaches. For Approach 1, location maps and the CDF plots are generated for two different frequencies. For Approach 2, only the location map is included. In our preliminary results in this paper,  we consider a high-SNR regime, and ignore the effects of noise for the sake of simplicity.  We will explore the performance of the proposed techniques at different  SNRs in our future work.


\begin{figure*}[t!]
\vspace{-.5cm}
\centering
\begin{subfigure}{.45\textwidth}
  \centerline{\includegraphics[scale=0.45]{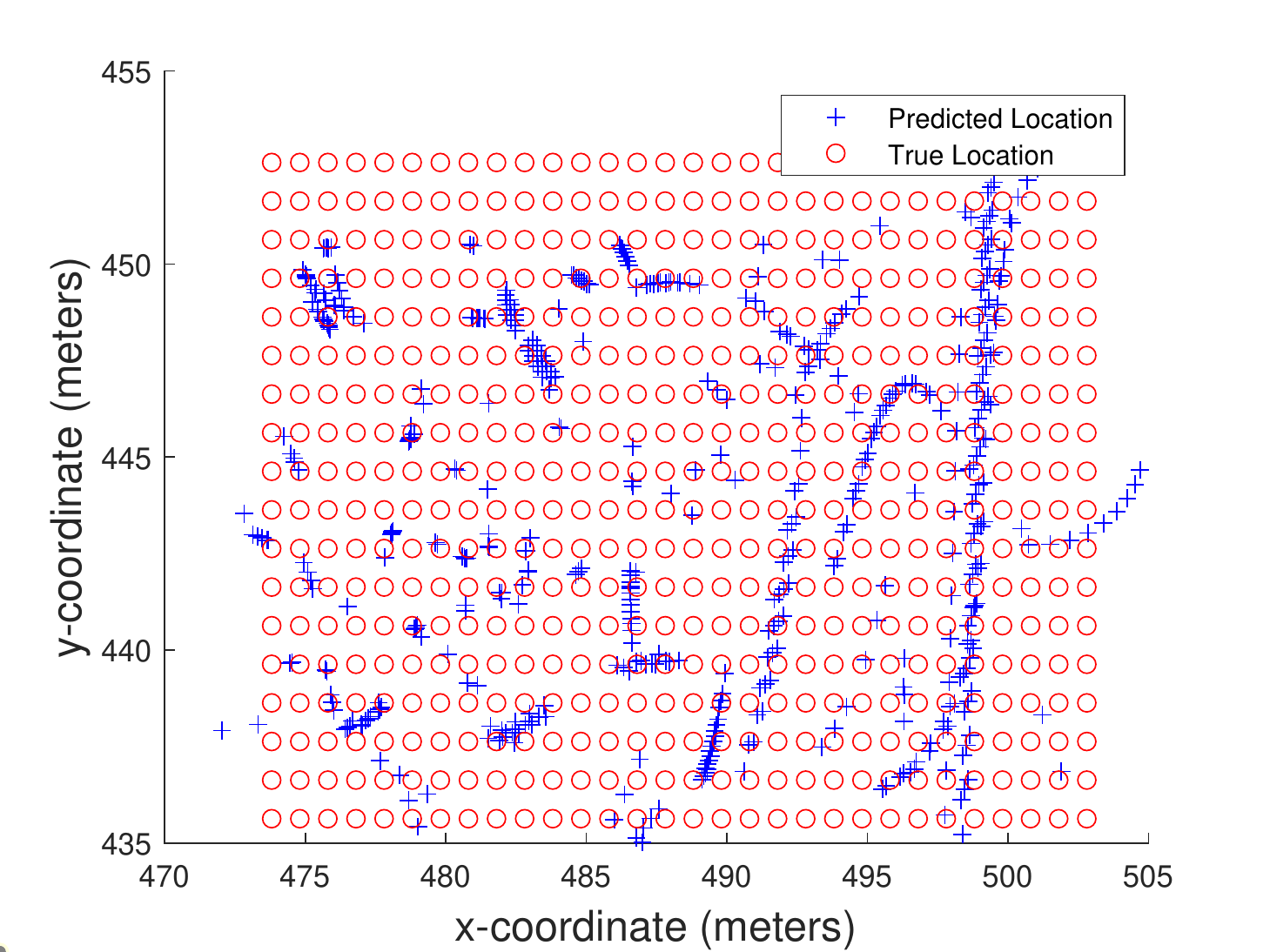} }
  \vspace{-.1cm}
  \caption{Actual and predicted location using only AOA.}
  \label{fig:sub-first}
\end{subfigure}
\begin{subfigure}{.45\textwidth}
  \centerline{\includegraphics[scale=0.45]{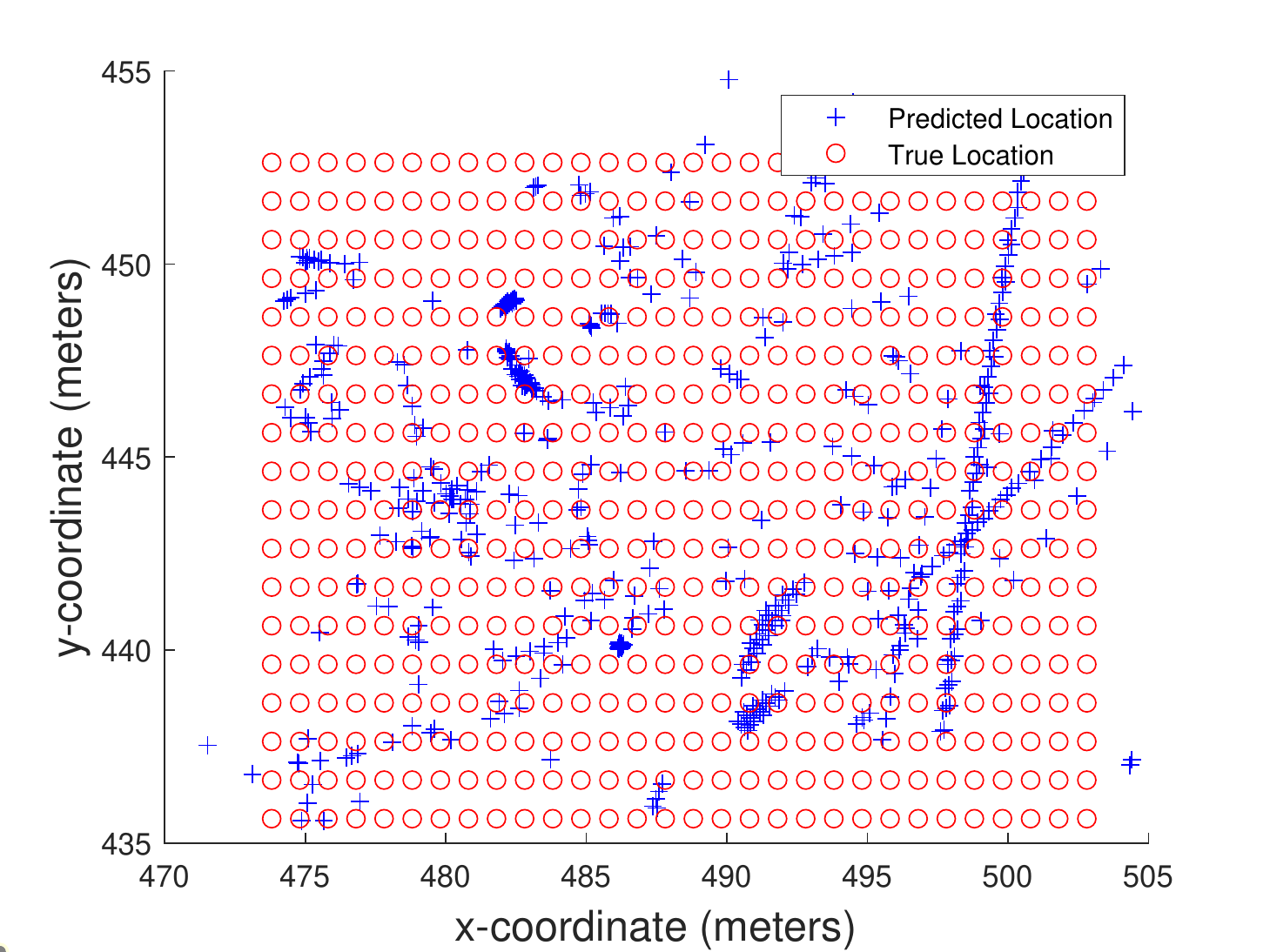}}
    \vspace{-.1cm}
  \caption{Actual and predicted location using AOA and RSS.}
  \label{fig:sub-second}
\end{subfigure}
\begin{subfigure}{.45\textwidth}
  \centerline{\includegraphics[scale=0.45]{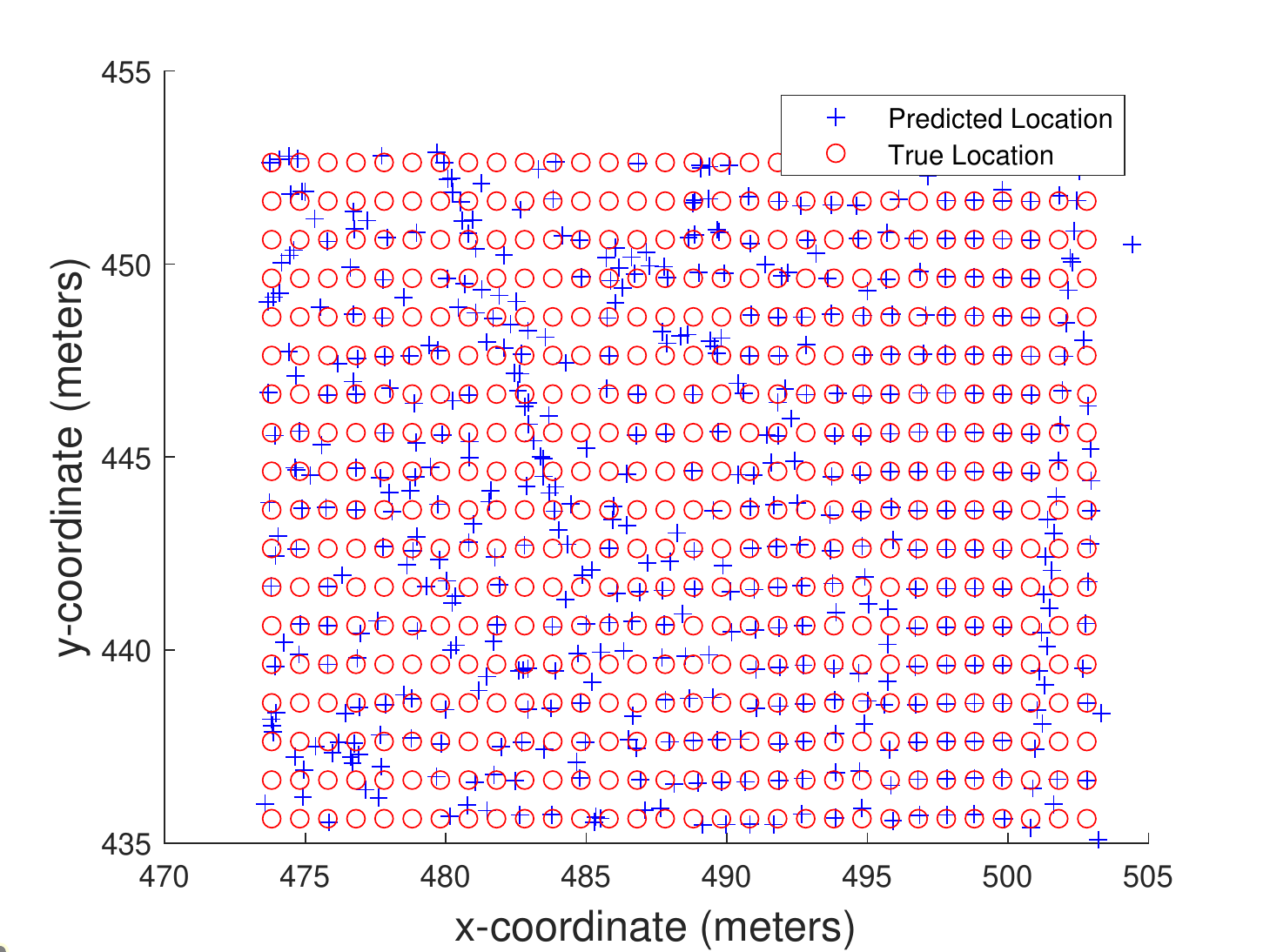} }
    \vspace{-.1cm}
  \caption{Actual and predicted location using AOA, RSS and TOA.}
  \label{fig:sub-third}
\end{subfigure}
\begin{subfigure}{.45\textwidth}
  \centerline{\includegraphics[scale=0.45]{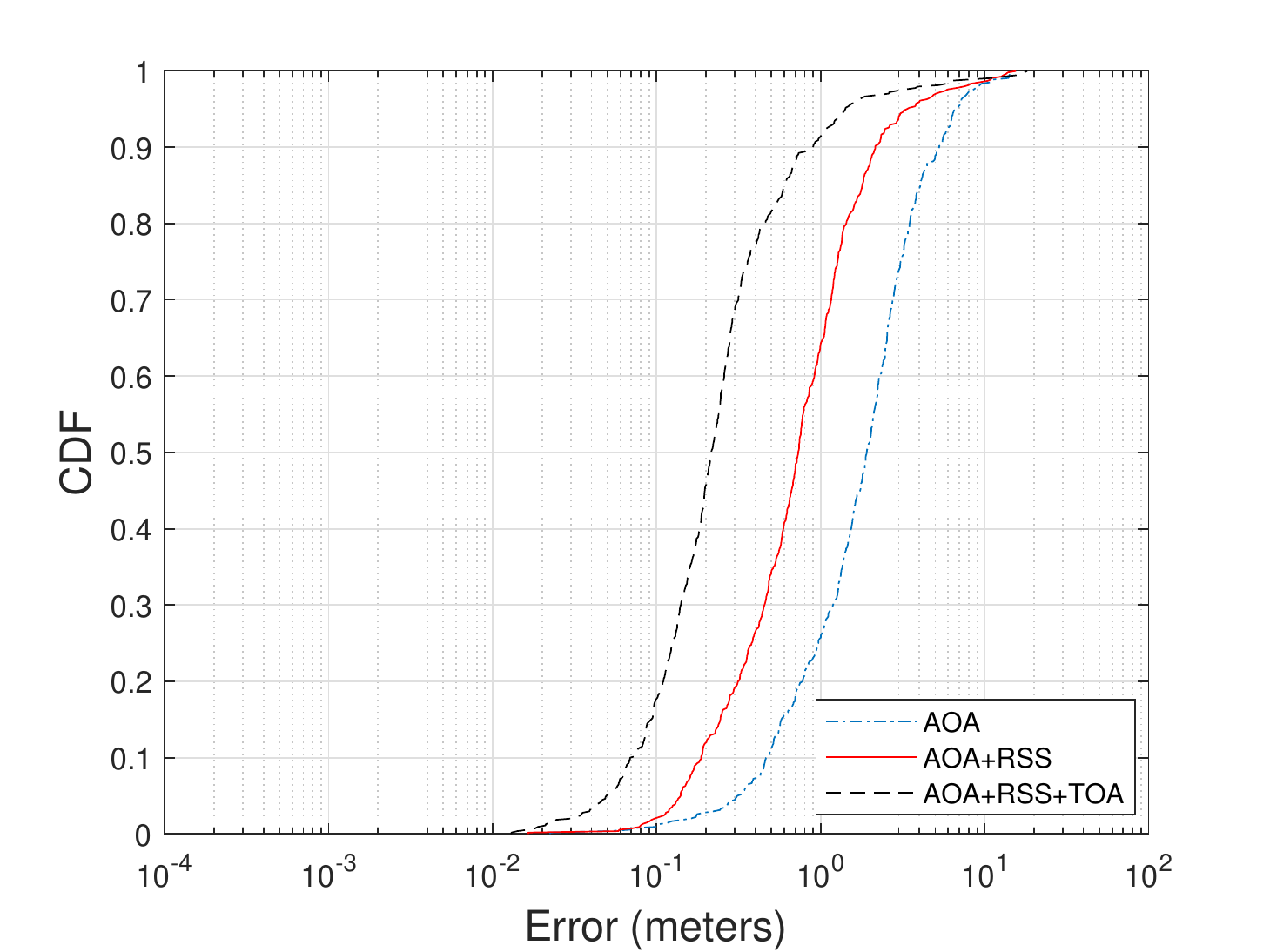} }
    \vspace{-.1cm}
\caption{CDF plot.}
  \label{fig:sub-fourth}
\end{subfigure}
\caption[1]{Location and CDF plot at 5~GHz (NLOS).}
\label{fig:5GHz_NLOS}
\end{figure*}

\begin{figure*}[h!]
\centering
\begin{subfigure}{.45\textwidth}
  \centerline{\includegraphics[scale=0.45]{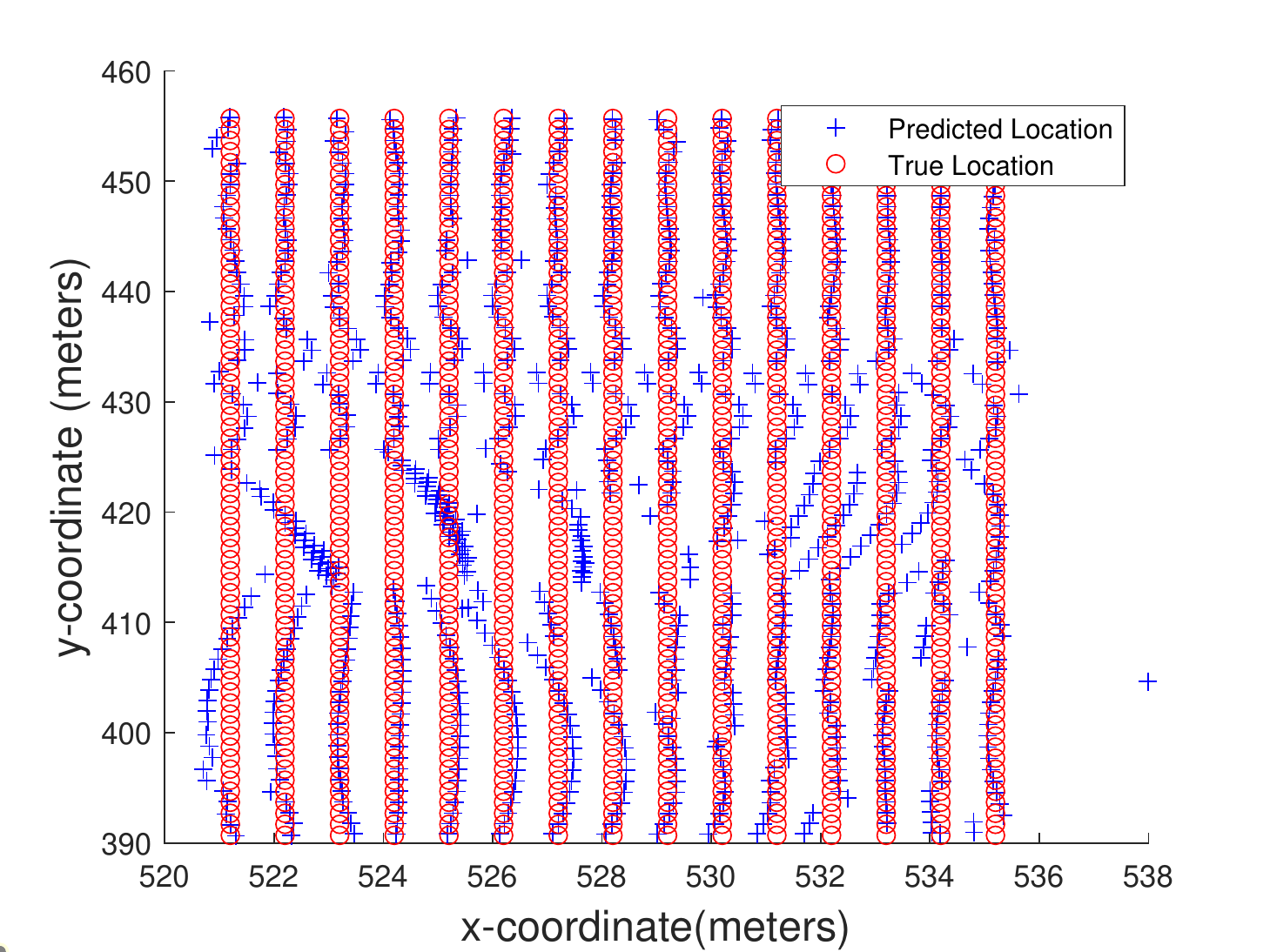} }
    \vspace{-.1cm}
  \caption{Actual and predicted location using only AOA.}
  \label{fig:sub-first}
\end{subfigure}
\begin{subfigure}{.45\textwidth}
  \centerline{\includegraphics[scale=0.45]{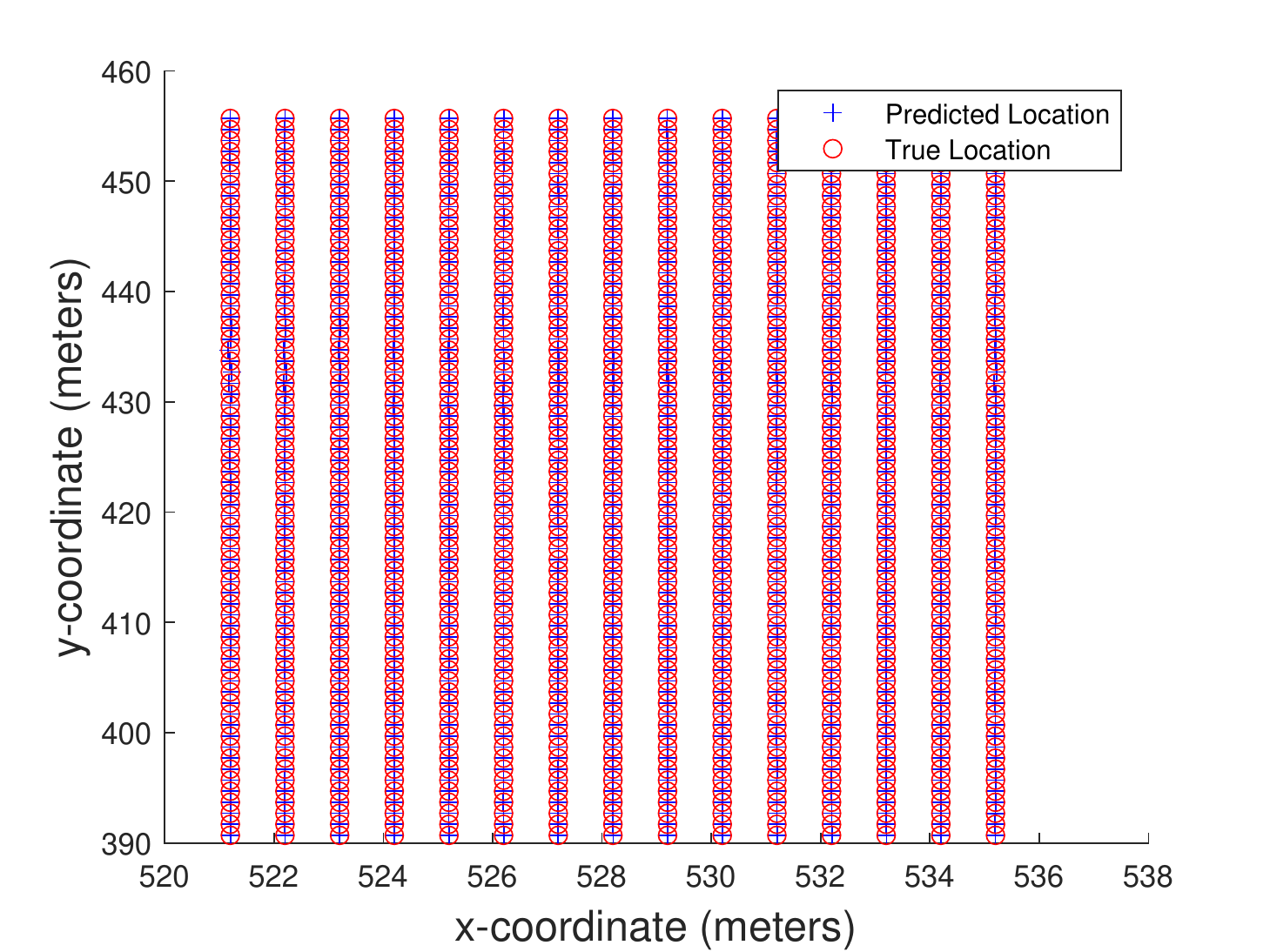}}
    \vspace{-.1cm}
  \caption{Actual and predicted location using AOA and RSS.}
  \label{fig:sub-second}
\end{subfigure}
\begin{subfigure}{.45\textwidth}
  \centerline{\includegraphics[scale=0.45]{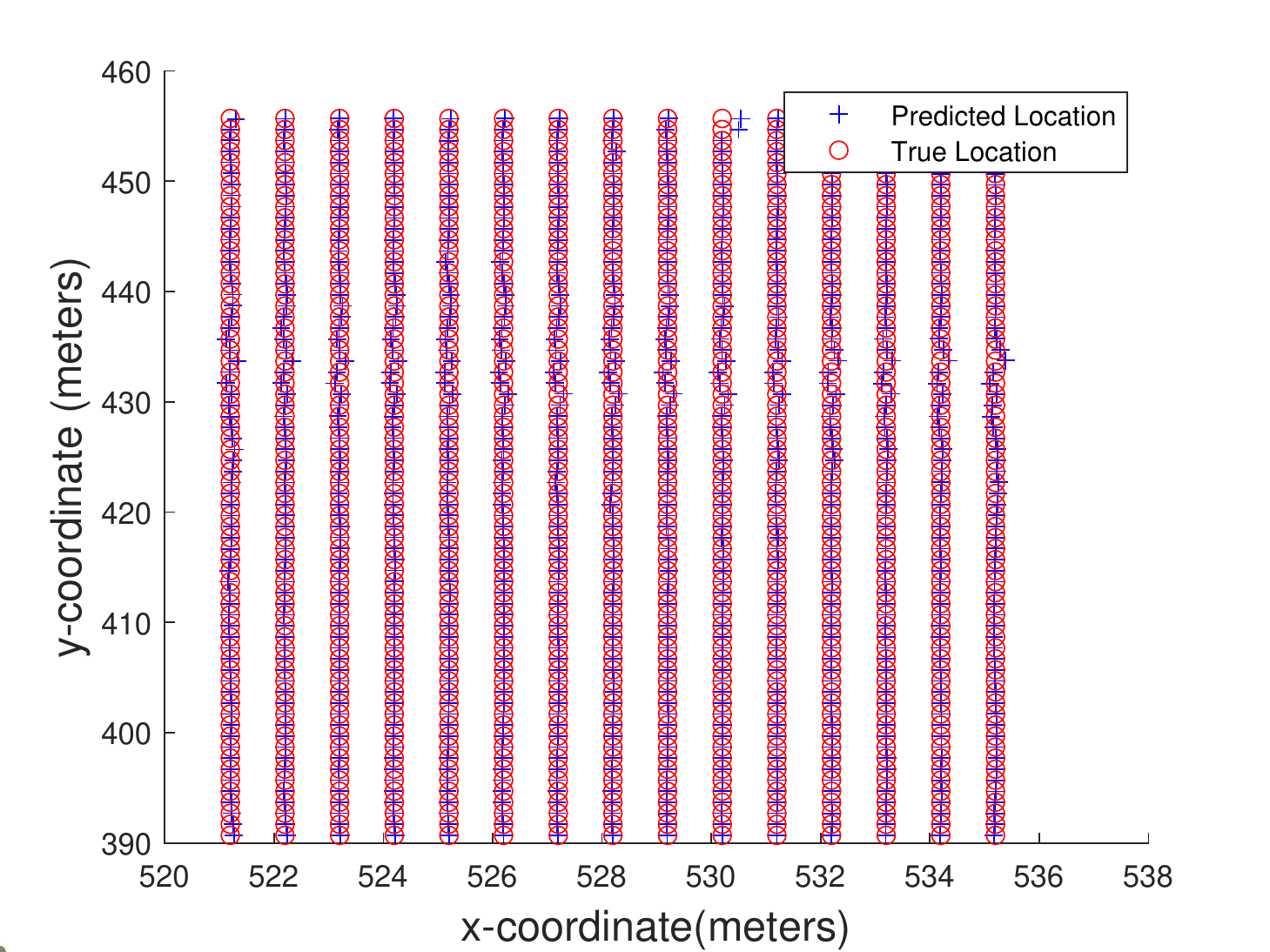} }
      \vspace{-.1cm}
  \caption{Actual and predicted location using AOA, RSS and TOA.}
  \label{fig:sub-third}
\end{subfigure}
\begin{subfigure}{.45\textwidth}
  \centerline{\includegraphics[scale=0.45]{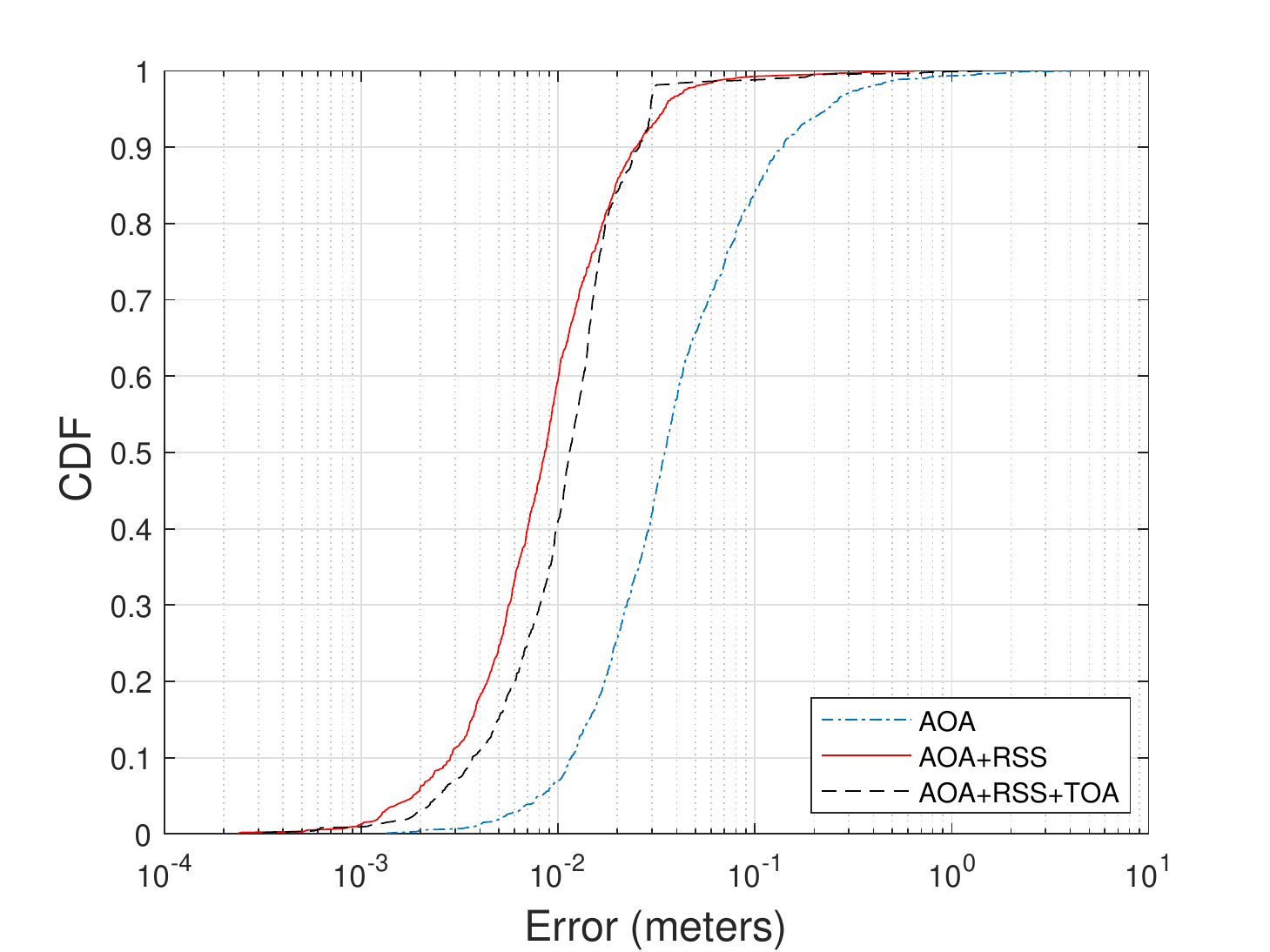} }
  \vspace{-0.1cm}
  \caption{CDF plot.}
  \label{fig:sub-fourth}
\end{subfigure}
\caption{Location and CDF plot at 28~GHz (LOS).}
\label{fig:28GHz_LOS}
\end{figure*}

\begin{figure*}[h!]
\centering
\begin{subfigure}{.45\textwidth}
  \centerline{\includegraphics[scale=0.5]{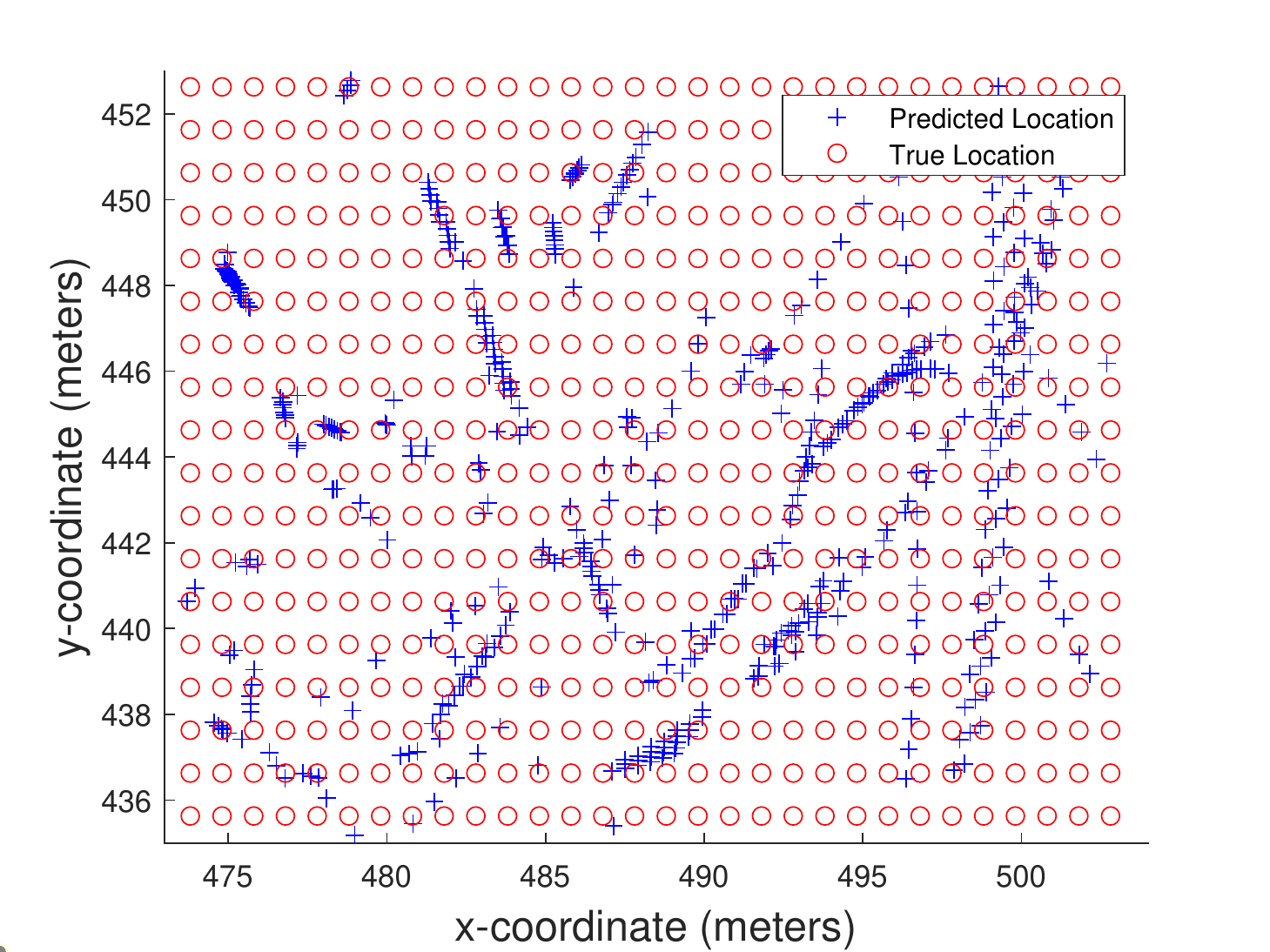} }
  \caption{Actual and predicted location using only AOA.}
  \label{fig:sub-first}
\end{subfigure}
\begin{subfigure}{.45\textwidth}
  \centerline{\includegraphics[scale=0.5]{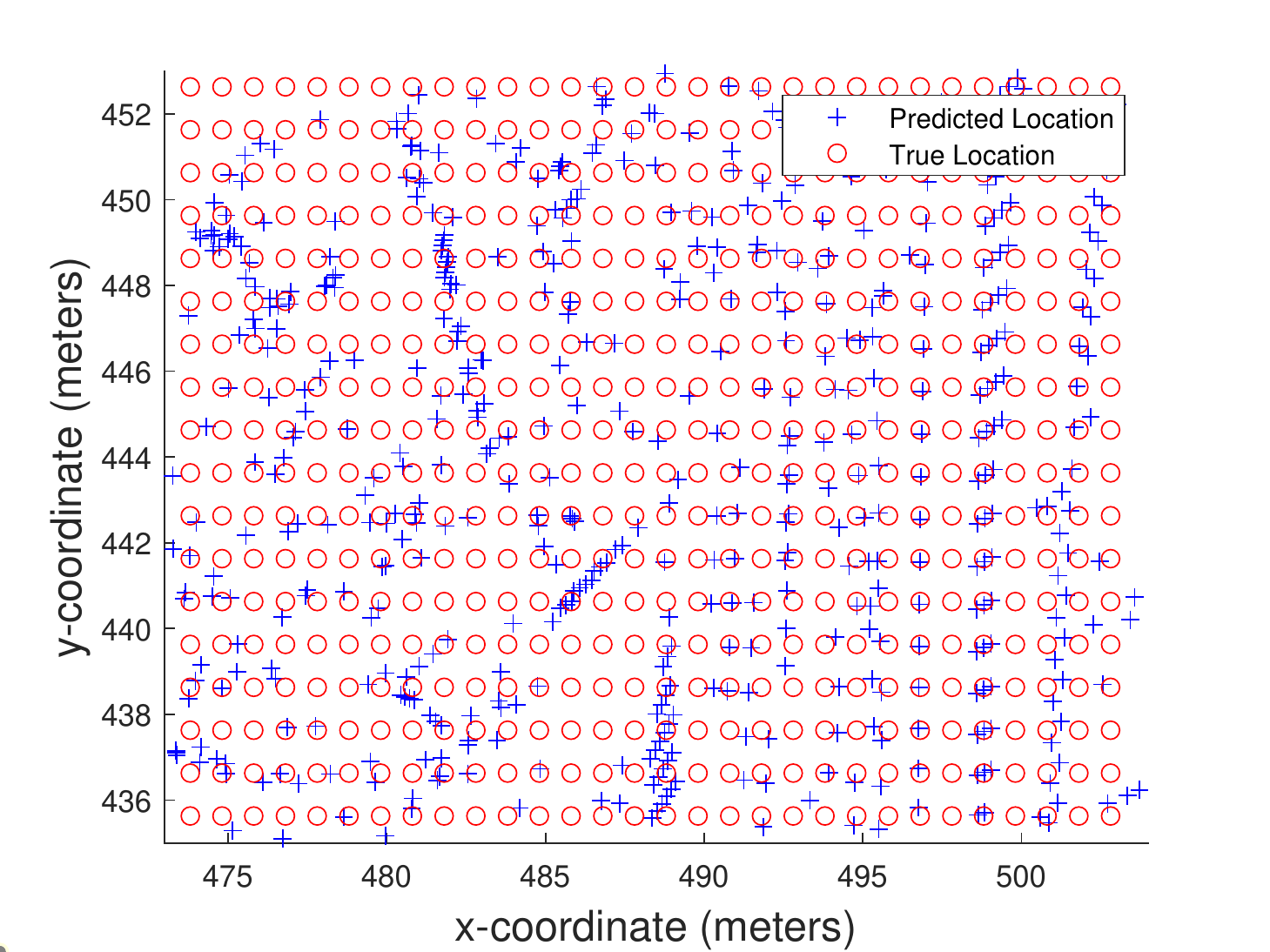}}
  \caption{Actual and predicted location using AOA and RSS.}
  \label{fig:sub-second}
\end{subfigure}
\begin{subfigure}{.45\textwidth}
  \centerline{\includegraphics[scale=0.5]{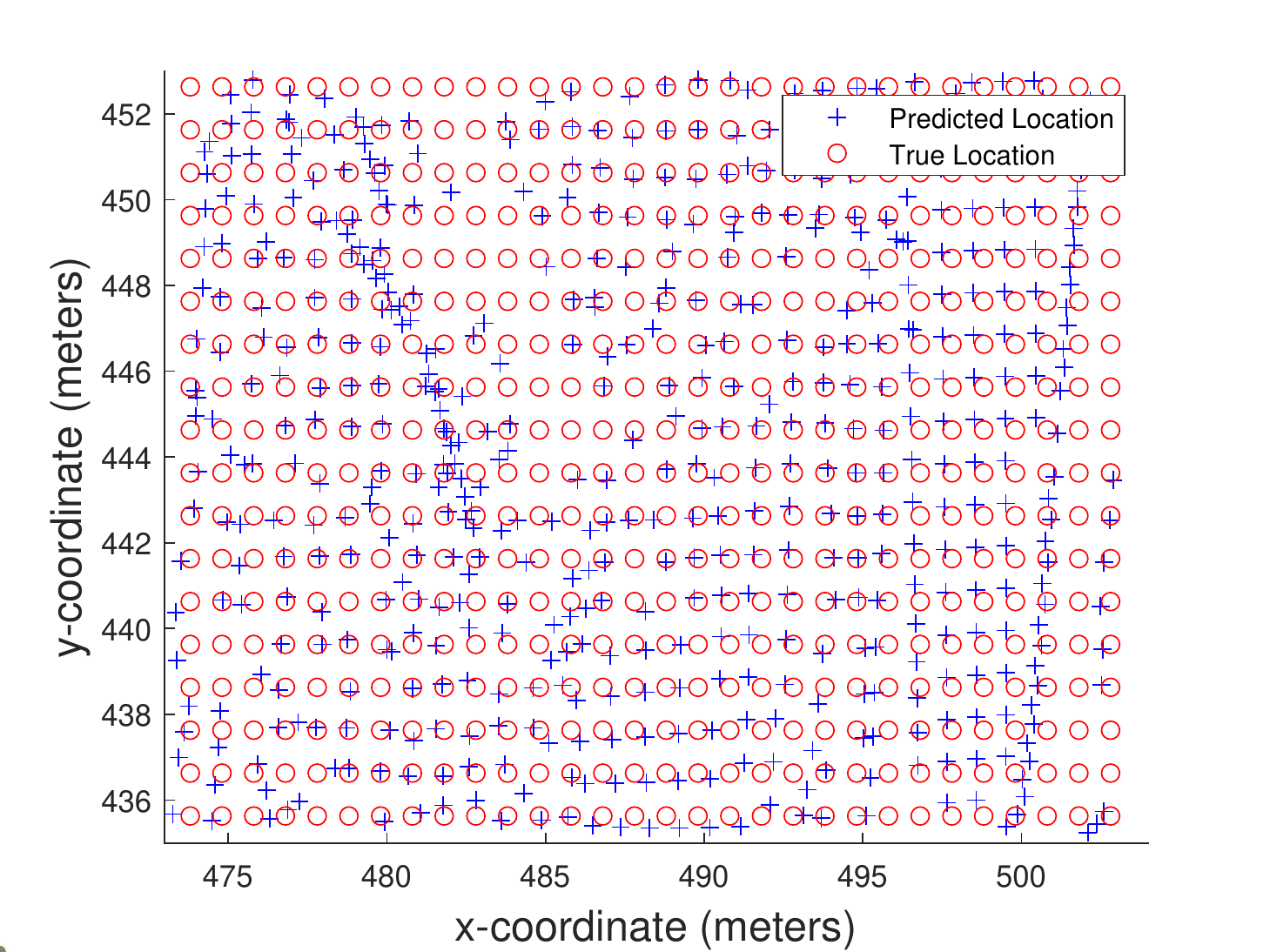} }
  \caption{Actual and predicted location using AOA, RSS and TOA.}
  \label{fig:sub-third}
\end{subfigure}
\begin{subfigure}{.45\textwidth}
  \centerline{\includegraphics[scale=0.5]{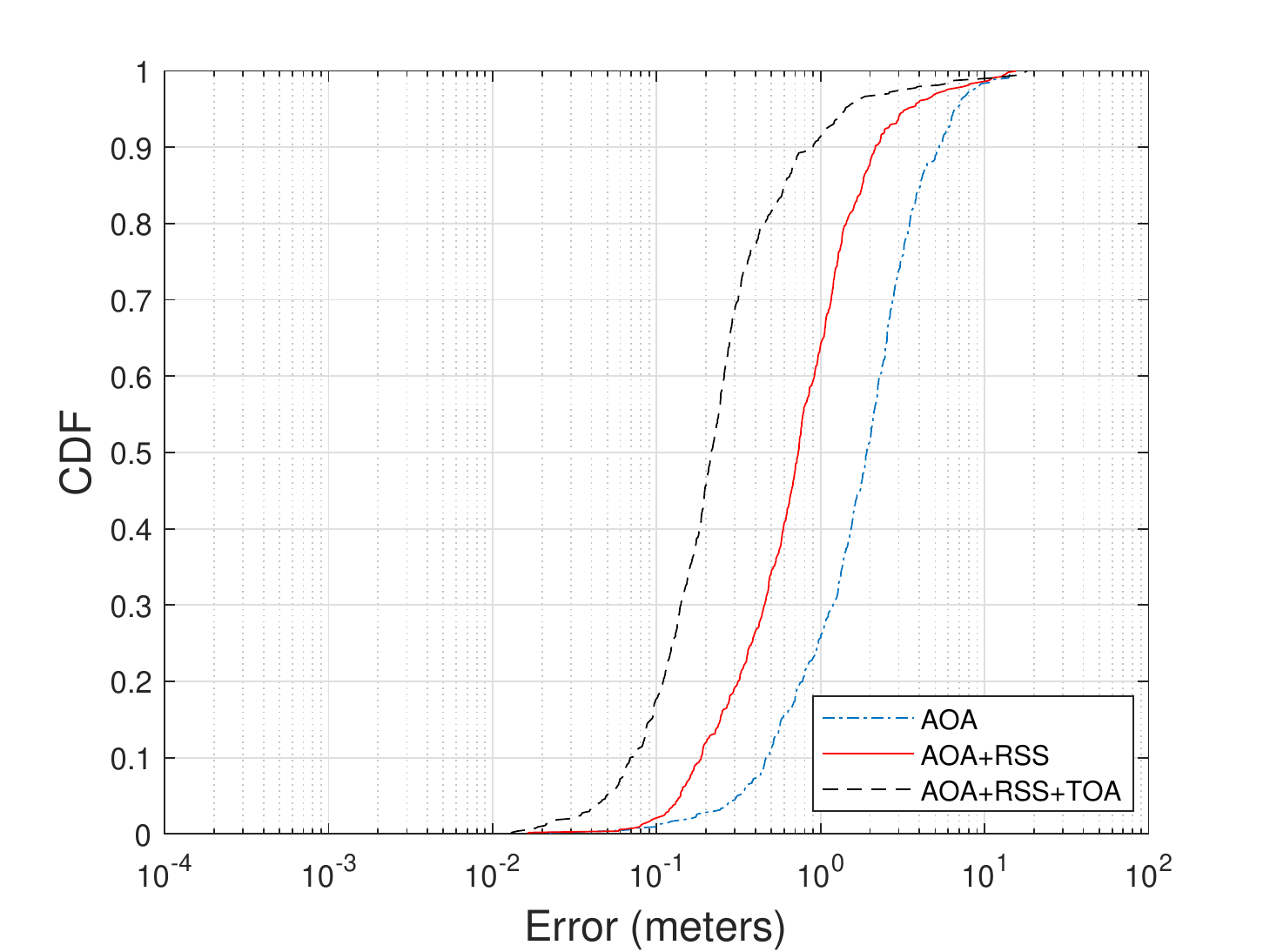} }
  \caption{CDF plot.}
  \label{fig:sub-fourth}
\end{subfigure}
\caption{Location and CDF plot at 28~GHz (NLOS).}
\label{fig:28GHz_NLOS}
\end{figure*}

\subsection{Bayesian Optimization Results}

The optimization procedure starts with a random combination of predefined range of hyper-parameters given in Table \ref{tab:T_FI_0}. The function 'bayseopt' calculates the cost function and accordingly chooses the next combination. We used mean squared error as the loss function and calculated the the total cost adding up all loss values as \eqref{Eq2}. After trying 30 different combinations, it picks the best combination to train the actual DNN. 
Table \ref{tab:T_FI} illustrates the best hyper-parameters obtained for different experiments. It is obvious from the table that the choice of hyper-parameters depends on the combination of input, scenarios, and frequency.

\subsection{Approach 1 Results}





Fig. \ref{fig:5GHz_LOS} (a), (b), (c) give the location mappings for LOS users (Receiver grid 2) at 5~GHz. At this scenario, considering only AOA gives the worst results. When RSS is given as input along with AOA, the accuracy of the model increases significantly. Adding TOA with the previous two features improves the model but not as much as it does when RSS is added. This is because RSS and TOA are somewhat correlated. For a user far from the BS, TOA will be greater, and RSS will be lower, and vice versa. Hence, in Fig. \ref{fig:5GHz_LOS} (d), the CDF for AOA + RSS and AOA+RSS+TOA curves are almost overlapping. On the other hand, the error is less than 0.1 for 90$\%$ of the users,  which is quite acceptable.

Fig. \ref{fig:5GHz_NLOS} (a), (b), (c) give the location mappings for NLOS users (receiver grid 3) at 5~GHz\footnote{Few predicted points that are outliers and significantly outside of the receiver grid have been ignored for the location maps for better visualization, but are included in the CDF plots.}. For NLOS case, the location mapping is not as accurate as it is for LOS scenario. In Fig. \ref{fig:5GHz_NLOS} (a), we observe that when only considering AOA, the error can be a maximum of 10~meters. However, as we increase the number of features i.e. adding RSS, and TOA, for 90$\%$ of the users, the error becomes less than 1 meter. For NLOS scenario at this frequency, the combination of RSS, TOA, and AOA gives better results than the combination of RSS and AOA, since more than one MPCs are considered. Adding an extra degree of freedom improves the results.

Fig. \ref{fig:28GHz_LOS} (a), (b), (c) give the LOS location mappings at 28~GHz. The trend of the results is similar to 5~GHz case. This time, adding TOA to the model as the third parameter does not increase the accuracy much. In Fig. \ref{fig:28GHz_LOS} (d), the CDF plot shows that almost all users are having an error of less than 0.1 meter when all three channel parameters are used. As expected, in the case of NLOS users at this frequency, the results are not as good as the LOS case. The model gives the best estimations when all three parameters are considered. 90$\%$ of the users are having an error of less than 1~meter in that case. 

Comparing the CDF curves for LOS and NLOS scenarios for both frequencies, it is observed that the curve is flatter in NLOS cases because the variation in error is higher. The performance of the DNN model towards NLOS points can be improved using more MPCs.

Comparing the results of 5~GHz and 28~GHz, one can observe the difference in the DNN performance. According to \cite{angular_correlation}, it is expected to have a significant angular congruency between different frequency bands. In other words, the AOA and the AOD is expected to be close for the corresponding MPCs in two bands. Intuitively, both the 5~GHz and 28~GHz cases should yield similar results especially for the DNN based on AOAs only. However, our results 
do not align with this intuition. 

One possible explanation for this could be the Bayesian optimization procedure, which starts the optimization with randomly chosen hyper-parameters that can be different in each training, even with the same inputs. As a result, the optimized final hyper-parameters can be different. Second, the angular congruency may not hold at 5~GHz and 28~GHz. In our ray tracing simulations, we limited the number of features to 3, and these are based on the 3 most dominant MPCs arranged in descending order with respect to their powers. Thus, even though there exist significant angular congruency across the bands, the 3 chosen MPCs based on the dominant power might have mismatch in the AOAs. One explanation to this could be that, the diffraction gets less effective as the frequency increases, therefore it is more likely for a diffracted signal to be more powerful than  reflected ones at 5~GHz band. We also see that 28~GHz frequency works better than sub-6~GHz band in NLOS scenarios and performance of the models do not differ much in LOS cases. We think that the resolution of the model should be better with smaller wavelenghts, thus it is expected to have more accurate results at mmWave frequencies. However, in LOS cases, information provided by the existing features are representative enough to estimate the location of the UEs such that the difference in resolution cannot be observed.

\subsection{Approach 2 Results}

\begin{figure}[t]
    \centerline{
    \includegraphics[scale=0.6]{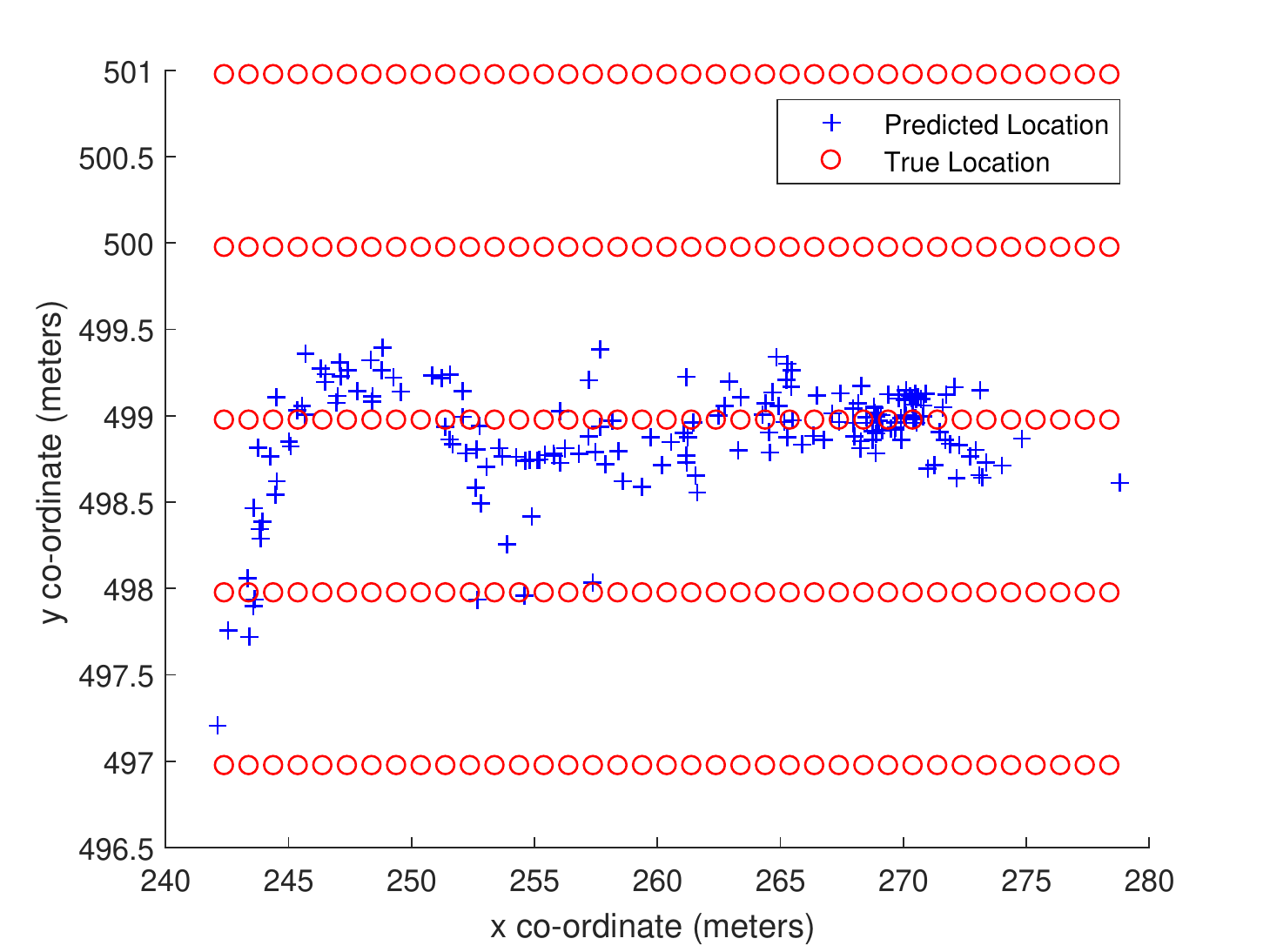}}
    \caption{Actual and predicted locations for approach 2 Deep-MIMO dataset.}
    \label{fig:Eta_Vs_Snap}
\end{figure}

When the DNN is trained using a channel vector response, it cannot predict the location as expected (Fig. \ref{fig:Eta_Vs_Snap}). The actual spacing between the users is about 1 meter. Since the users are closely located, there is not much difference in the channel response. Hence, the DNN fails to train itself properly. This problem can be solved by the feature transform technique used in \cite{studer2018channel}. Using this technique, the input features can be sparsely distributed so that the DNN can easily differentiate between two users, in terms of their channel response. The implementation of a feature transform is considered as a future work.

\section{Conclusion and Future work}
This work gives an idea of implementing DNN in localization providing a comparison between mmWave and sub-6~GHz bands, LOS and NLOS scenarios as well as two different approaches, considering a high-SNR regime. Using three channel parameters for fixed number of (three) MPCs that are used in the DNN, the location of a UE can be predicted with very high accuracy. 
On the other hand, the presumption of using only three MPCs can be overcome using channel response with feature transform as the input. 

The DNN is trained using supervised learning techniques in this work, which is possible for synthetic data. Getting sufficient real data, which is costly, could make it possible to train the DNN using semi-supervised or unsupervised learning techniques. In that case, derived features could be used rather than the raw features. We considered a static scenario in both of the approaches. In order to make the simulation environment more realistic, a more dynamic scenario should be considered, which will be tackled as future work along with performance evaluation at various different SNRs. We also plan to implement this technique to a 3D scenario to localize aerial users such as unmanned aerial vehicles (UAVs) in urban, sub-urban and rural areas.

\section{Acknowledgment}
This work was supported in part by NSF under the grant number CNS-1814727, and in part by INL Laboratory Directed
Research \& Development (LDRD) Program under DOE Idaho Operations Office Contract DE-AC07-05ID14517.
\bibliographystyle{IEEEtran}
\bibliography{IEEEabrv,papers}

\end{document}